\documentclass[preprint,12pt]{aastex}

\shorttitle{Theoretical Lithium Depletion Boundary}
\shortauthors{Burke, Pinsonneault, \& Sills}

\begin{document}

\title{Theoretical Examination of the Lithium Depletion Boundary}
\author{Christopher J. Burke and Marc H. Pinsonneault}
\affil{Astronomy Department, The Ohio State University, 140 W. 18th Ave., Columbus, OH 43210;\\cjburke@astronomy.ohio-state.edu, pinsono@astronomy.ohio-state.edu}

\and

\author{Alison Sills}
\affil{Department of Physics and Astronomy, McMaster University, 1280 Main Street West, Hamilton, Ontario, Canada, L8S 4M1;\\asills@physics.mcmaster.ca}

\begin{abstract}

We explore the sensitivity in open cluster ages obtained by the
lithium depletion boundary (LDB) technique to the stellar model input
physics.  The LDB age technique is limited to open clusters with ages
ranging from 20 to 200 Myr.  Effective 1-$\sigma$ errors in the LDB
technique due to uncertain input physics are roughly 3\% at the oldest
age increasing to 8\% at the youngest age.  Bolometric correction
uncertainties add an additional 10 to 6\% error to the LDB age
technique for old and young clusters, respectively.  Rotation rates
matching the observed fastest rotators in the Pleiades affect LDB ages
by less than 2\%.  The range of rotation rates in an open cluster are
expected to ``smear'' the LDB location by only 0.02 mag for a Pleiades
age cluster increasing to 0.06 mag for a 20 Myr cluster.  Thus, the
observational error of locating the LDB ($\sim$7-10\%) and the
bolometric correction uncertainty currently dominate the error in LDB
ages.  For our base case, we formally derive a LDB age of 148$\pm$ 19
Myr for the Pleiades, where the error includes 8, 3, and 9\%
contributions from observational, theoretical, and bolometric
correction sources, respectively.  A maximally plausible 0.3 magnitude
shift in the I-band bolometric correction to reconcile main sequence
isochrone fits with the observed (V-I) color for the low mass Pleiades
members results in an age of 126$\pm$ 11 Myr, where the error includes
observational and theoretical errors only.  Upper
main-sequence-fitting ages that do not include convective core
overshoot for the Pleiades ($\sim$75 Myr) are ruled out by the LDB age
technique.
\end{abstract}


\section{Introduction} Accurate ages for nearby stellar clusters are
the basic observational templates that constrain the formation history
of our Galaxy and the Universe.  Unfortunately, theoretical isochrones
fit to the upper main sequence provide absolute ages for the most
thoroughly studied young open clusters that are still uncertain by a
factor of two \citep{sta00}.  The uncertainty mainly results from the
dependence of the lifetime of massive stars on the size of their
convective cores.  Convective core overshoot brings fresh
hydrogen-rich material to the core, extending the stellar lifetime on
the main sequence.  As a result, cluster age estimates are quite
sensitive to the degree of convective core overshoot in the
theoretical calculations.  There is evidence that the inclusion of
convective core overshoot results in an improved agreement between
theoretical evolution rates and number counts in the Hertzsprung gap,
and also an improved fit to the width of the main sequence turnoff in
open clusters \citep{and90,dem94}.  Thus, theoretical calculations
that include convective core overshoot result in an older absolute age
scale relative to theoretical calculations that do not.  Calibrating
the older convective core overshoot age scale requires an independent
technique for determining stellar cluster ages.  Lithium depletion age
dating is one possibility, and in this paper we discuss the
reliability of this technique.

The lithium depletion boundary (LDB) technique is an independent
method to determine the age of open clusters with ages that range from
20 to 200 Myr.  Proton reactions destroy Li$^{7}$ near a destruction
temperature, T$_{D}\sim 2.5\times 10^{6}$ K, easily obtained under
stellar conditions.  In general, the presence or absence of
photospheric lithium determines whether sufficient time has passed for
a majority of the stellar material to reach depths in the star where
T=T$_{D}$.  For stars that are fully convective during
pre-main-sequence contraction (M$\la 0.4$ M$_{\odot}$), the convective
overturn timescale is much less than the evolutionary timescale, and
the entire lithium content is rapidly destroyed when the stellar core
temperature, T$_{C}$, reaches T$_{D}$.  Since the rate at which the
core temperature increases to the destruction temperature is a strong
function of stellar mass, spectral observations of Li$^{7}$ in fully
convective stars during pre-main-sequence contraction is an accurate
age diagnostic.  Higher mass stars reach the condition T$_{C}=$T$_{D}$
at an earlier age and higher luminosity than lower mass stars, and the
lithium abundance decreases by a factor of 100 over a very narrow
luminosity.

LDB ages have been obtained for several young open clusters: the
Pleiades, Alpha Per, IC 2391, \& NGC 2547
\citep{sta98,sta99,bar99,oli03}.  In an independent analysis,
\citet{jef00} verified the conclusions of the previous open cluster
studies that the LDB ages are $\sim$1.6 times older than upper
main-sequence-fitting ages without convective core overshoot.
\citet{jef00} find that the observational uncertainties (locating the
LDB, photometric calibrations, distance, and reddening) are larger
than the theoretical uncertainties (choice of model and bolometric
corrections) for the Pleiades ($\sigma_{obs}\sim 16\%$;
$\sigma_{thy}\sim 9\%$, respectively).  However, \citet{jef00} point
out that the theoretical errors may not be complete, since common
assumptions made by several independent theoretical calculations used
in this study may hide additional systematics.  They also point out
 that non-negligible errors may be associated with the bolometric
 corrections.

\citet{bil97} provide the most detailed quantitative estimate of the
uncertainties in LDB calculations.  The \citet{bil97}, analytical
lithium depletion model consists of a fully convective polytrope
contracting at constant effective temperature.  A comparison of their
analytical model to detailed numerical calculations results in maximum
differences in the LDB ages of 25\% from \citet{cha96} and 15\% from
\citet{dan94}, demonstrating the robustness of the LDB technique to
uncertain input physics.  \citet{dan97} comment on the LDB ages being
``practically independent on the details of the models including the
opacity''.  However, they do not go into quantitative detail.

In this paper we determine the size of errors in the LDB technique
that result from theoretical errors in the lithium depletion
calculations and uncertainties in the bolometric corrections.  We can
then determine the dominant error source in the LDB age technique and
confirm whether the discrepancy between the upper
main-sequence-fitting ages and LDB ages is real.  Employing the Yale
Rotating Stellar Evolution Code (YREC), we compare the lithium
depletion calculations using up-to-date input physics to models using
older generation equations of state, opacities, and atmospheres as
well as variations in the mixing length and metallicity.  The
difference in LDB ages that results from employing older generation
input physics allows us to ascertain the expected deviations of the
LDB ages from up-to-date input physics.  Additionally, we quantify the
impact of stellar rotation and various bolometric correction
determinations on LDB ages.

In Section~\ref{fiducial} we briefly describe the YREC stellar model
and determine our reference LDB-luminosity-age relation
obtained with the latest input physics.  Section~\ref{interr} examines
the robustness of the LDB to variations in the input physics.  Then,
we discuss the error budget in the LDB ages due to input physics
uncertainties in Section~\ref{interrbudget}.  In Section~\ref{exterr}
we compare the reference LDB calculations to other calculations in the
literature.  The error budget in the LDB ages derived from the
comparisons with other calculations in the literature and the final
adopted theoretical error in the LDB ages is discussed in
Section~\ref{exterrbudget}.  Section~\ref{bolosec} is devoted to the
significant uncertainties in the I-band bolometric correction.  In
Section~\ref{conclude} we summarize our results and derive new LDB
ages with improved errors for the open clusters with previous LDB age
determinations.

\section{Reference Model} \label{fiducial} In this section we describe
the lithium depletion calculations and the input physics that define
our reference LDB-luminosity-age relation.  This reference relation
serves as the standard to quantify the uncertainty in LDB ages.  We
use the Yale Rotation Evolution Code (YREC) \citep{sil00} for all
stellar model calculations.  For the reference calculation, we select
standard, non-rotating, solar-calibrated stellar models with masses
that range from 0.065 to 0.30 M$_{\odot}$.  The upper mass limit
ensures full convection during lithium depletion and minimizes the
impact of initial conditions.  Through accretion and mass loss, a star
interacts with its surroundings during the earliest phases of its
life.  However, any structural changes to the star that may occur due
to this complex interaction with its surroundings is erased after
several Kelvin-Helmholtz time scales.  The Kelvin-Helmholtz time scale
for our M=0.3 M$_{\odot}$ starting model is one-tenth of the time for
lithium depletion, making these complications negligible.  Stars with
M$\leq$0.06 M$_{\odot}$ never reach temperatures high enough to
destroy lithium completely \citep{dan94}, and our reference
calculations confirm this fact.  The above mass limits constrain the
validity of the LDB age technique to stellar populations with ages
between 20 and 200 Myr.  For clusters with ages less than 20 Myr, the
uncertain initial conditions, deuterium burning, and adopted zero-age
(which all occur on timescales of a few 10$^{6}$ years) become an
increasing fraction of the LDB age.  For clusters with ages older than
200 Myr, lithium is depleted over a larger width of luminosity, and
lacks a clearly defined LDB.

Adopting a solar metallicity, Z=0.0188, and the heavy element mix of
\citet{gre93}, we determine the ratio of mixing length parameter to
pressure scale height and helium abundance by calibrating to the solar
data, resulting in $\alpha\equiv l_{m}/H_{p}=1.75$ and Y=0.27,
respectively.  We use the OPAL equation of state from \citet{rog96}
where it is available and the equation of state of \citet{sau95}
otherwise.  We investigate the effect of the latest OPAL 2001 equation
of state \citep{rog02} on the LDB ages in Section~\ref{eos}.  The low
temperature atmospheres of \citet{hau99} provide the pressure at
T=Teff, which we apply as the outer boundary condition.

\begin{figure}
\plotone{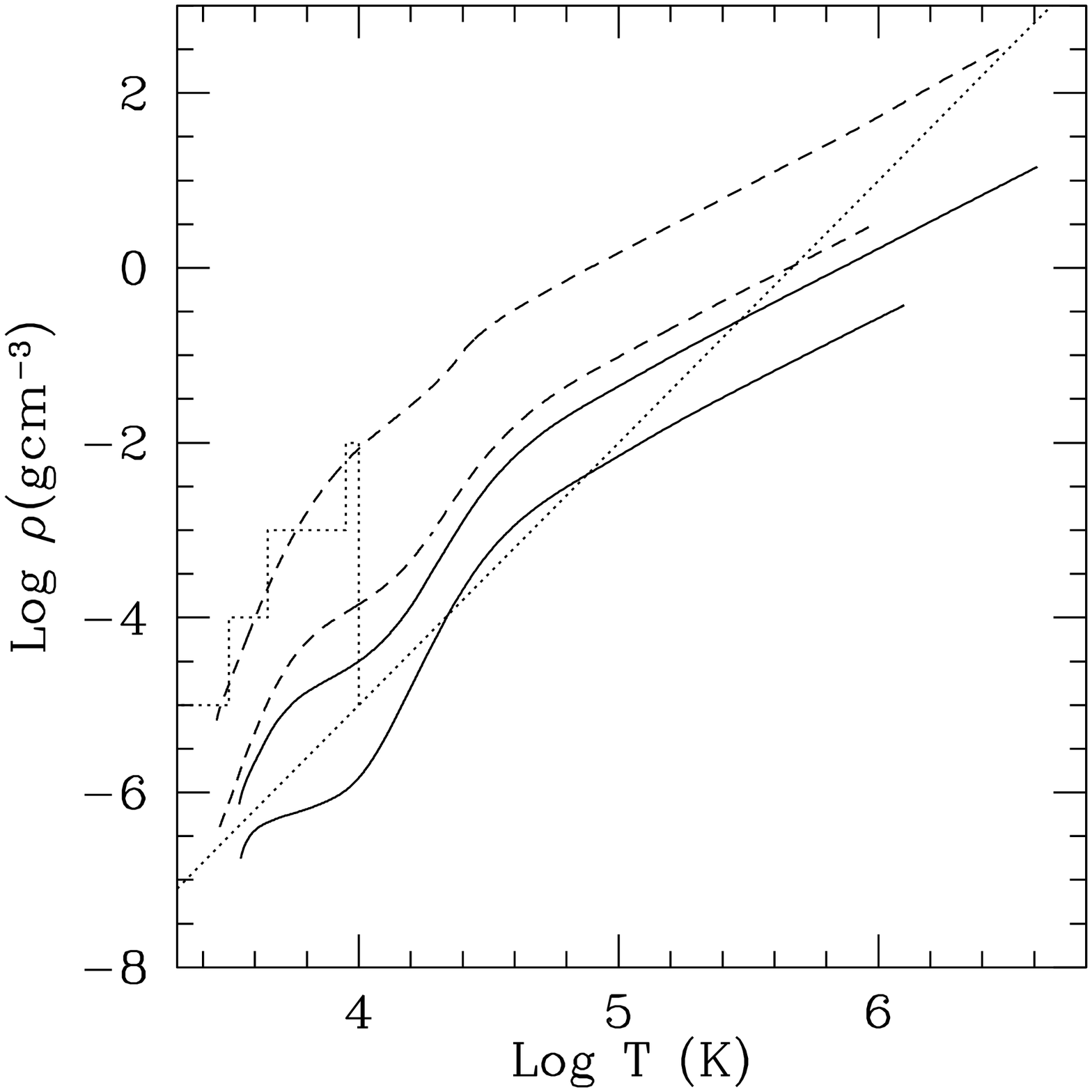}
\caption{Solid lines are the run of density as a function of
temperature for M=0.30 M$_{\odot}$ for the initial and lithium
depletion models, bottom and top respectively.  Dashed lines show the
run of density for the M=0.065 M$_{\odot}$ models.  The diagonal
dotted line represents the upper density boundary at
$R\equiv\rho/$T$^{3}_{6}=1$ for the \citet{ale94} low temperature and
\citet{igl96} high temperature opacities used in the reference
calculations.  The dotted line extending to higher densities shows the
\citet{ale83} molecular opacity table boundaries.}
\label{runrho}
\end{figure}

Physical conditions in low mass stars fall outside opacity tables
employed to model solar-type stars.  Figure~\ref{runrho} shows several
stellar density profiles as a function of temperature for the low mass
stellar regime applicable to this study.  The two solid lines trace
the run of density for the 0.3 M$_{\odot}$ reference models at the
initial and lithium depletion epochs; lower and higher density,
respectively.  The two dashed lines show the density evolution for the
0.065 M$_{\odot}$ reference model.  The dotted lines represent the
various opacity table boundaries.  The diagonal dotted line in
Figure~\ref{runrho} shows the upper density boundary at
R$\equiv\rho/$T$^{3}_{6}=1$ for the \citet{ale94} low temperature and
\citet{igl96} high temperature opacities used in the reference
calculations.  A linear ramp between 4.0$\la \log$(T)$\la$4.1 provides
a smooth transition between the two opacity tables.  For the reference
model, we supplement the \citet{ale94} low temperature opacity toward
higher densities with the molecular opacity table of \citet{ale83}.
In Figure~\ref{runrho}, the dotted line starting at $\log$(T)=4.0 with
an uneven upper-density boundary toward lower temperatures outlines
the \citet{ale83} molecular opacity table.  A linear ramp toward
higher densities over 1 dex starting at the R=1 boundary is used to
transition smoothly between the \citet{ale94} and \citet{ale83}
opacities.  A linear extrapolation in density provides the opacity for
densities higher than the table boundaries.

The OPAL opacity table \citep{igl96} and OPAL equation of state
\citep{rog96} share the same R=1 boundary shown in
Figure~\ref{runrho}.  For densities higher than the R=1 boundary, we
apply a linear ramp constant in density to transition to the
\citet{sau95} equation of state.  The ramp has a width of the final
two density values of the OPAL table.  The OPAL equation of state also has
a low-temperature boundary at $\log$(T)=3.75 K.  For temperatures
below the R=1 boundary, we transition to the \citet{sau95} equation of
state with a linear ramp starting at $\log$(T)=3.80 K.
 
We update the conductive opacities with the extensive calculations
from \citet{pot97} and \citet{pot99}.  Their conductive opacity
calculations are for a fully ionized single-ion species.  We follow
the procedure given in \citet{pot99} to calculate an arbitrary mixture
for the conductive opacity by expressing the electron-ion collision
frequency as a summation over the coulomb logarithm factors for the
ion species present.  We assume singly ionized O$^{16}$ for the entire
stellar metal content.  H and He are assumed fully ionized, valid
since in the unionized regime, $\log(\rho)\la 0.0$ and
$\log$(T)$\la 4.0$, conduction is inefficient and the radiative
opacity dominates.

The initial stellar models start at the deuterium-burning
birthline \citep{sta88}.  We obtain our deuterium birthline by starting
a stellar model high up on the Hyashi track and designate the
birthline when the core temperature reaches the same core temperature
when deuterium decreases by a factor of 100 in the calculations of
\citet{dan94} (see their Table 7).  Since deuterium burning completes
in $\lesssim 3\%$ of the time for lithium depletion, our results are
insensitive to the adopted initial conditions.

Using the above mentioned input physics, we calculate a reference
relation for the luminosity and age when lithium is depleted by a
factor of 100 from its initial value.  The lithium depletion ages are
systematically 15\% and 6\% younger at fixed luminosity for adopting a
lithium depletion by a factor of 2 and 10, respectively.  As we
quantify in the following sections, the systematic LDB age difference
associated with adopting a different lithium depletion factor is
similar in size to errors associated with the input physics
uncertainty.  Thus, comparisons with other lithium depletion
calculations require a common depletion factor.

\begin{figure}
\plotone{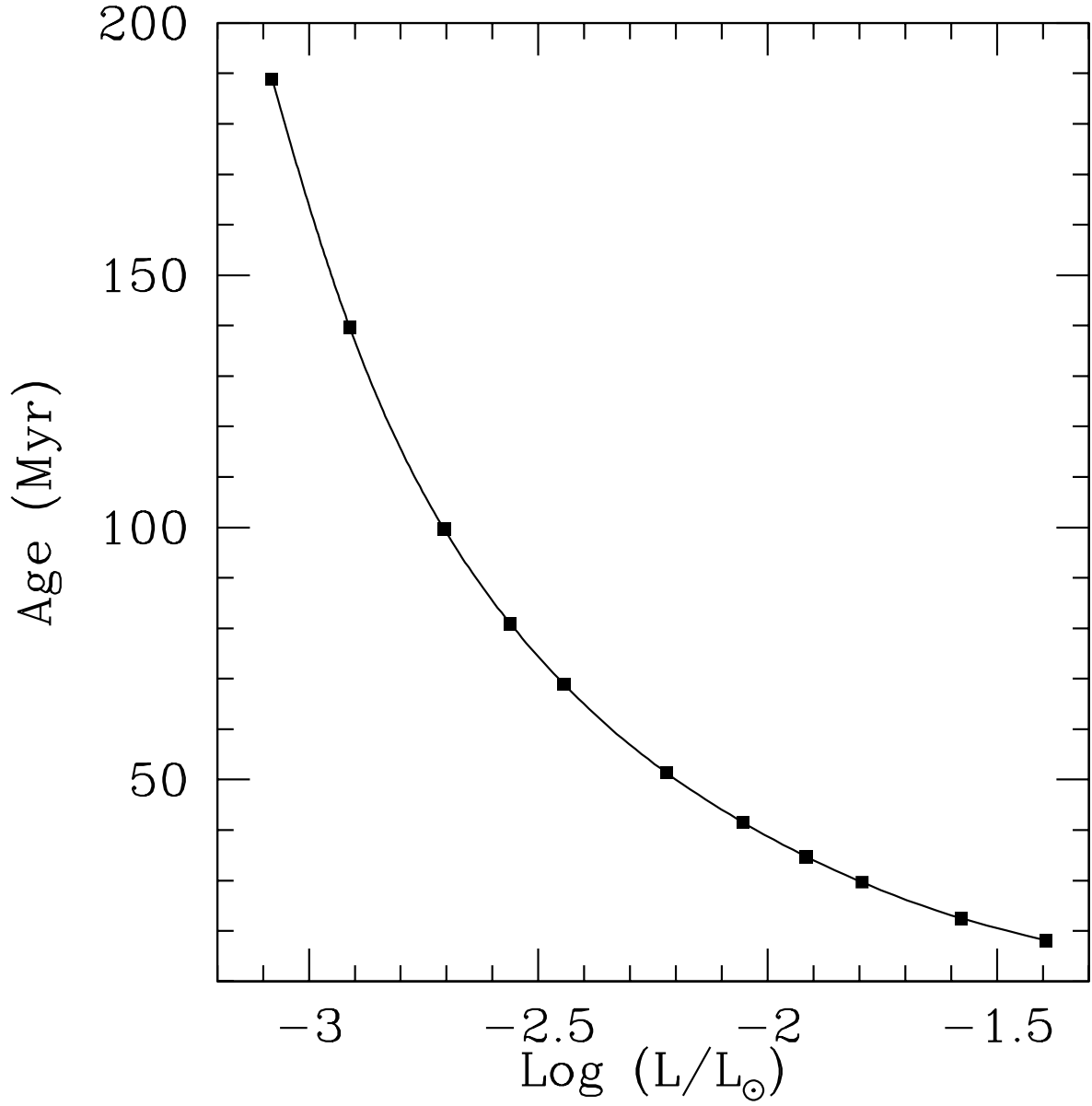}
\caption{Reference lithium depletion age as a function of luminosity.}
\label{refli7age}
\end{figure}

We find the LDB age is sensitive to numerical time resolution.
Insufficient time resolution results in systematically older lithium
depletion ages.  Halving the time resolution iteratively until the
lithium depletion age varies by less than 1\% ensures the
time-resolution numerical systematic error is negligible.  Doubling
the spatial resolution of our reference calculation results in no
impact on the derived LDB ages as a function of luminosity.  When
examining the sensitivity of the LDB ages to changes in the input
physics, all LDB age variations are given as deviations from this
reference model.  The reference LDB-luminosity-age relation is shown
in Figure~\ref{refli7age}.  The relation in tabular form is given in
Table~\ref{reftab}.

\section{Input Physics Uncertainties} \label{interr} In this section
we determine the error in the LDB-luminosity-age relation that results
from uncertainties in the model input physics.  We quantify the
theoretical uncertainty by varying the input physics of our reference
model.  The following subsections describe each physical input in
turn.

\subsection{Mixing Length} Our models employ the mixing length theory
for convective transport \citep{vit53}.  As in our reference models,
the mixing length parameter is commonly adjusted to fit the solar
data.  However, without a physical model for the mixing length
parameter, the solar calibrated model may not be applicable to stars
of a different mass \citep{lud99}.  Other studies \citep{bar98} adopt
the mixing length to be equal to the pressure scale height,
$\alpha=1.0$, for the interior calculation in order to match the
$\alpha=1.0$ employed in the atmosphere \citep{hau99}.

The temperature gradient only becomes sensitive to the adopted mixing
length parameter in the superadiabatic zone of the low density
envelope.  The deep interior of the star has vigorous convection
characterized by the temperature gradient, $\nabla = \nabla_{ad}$
\citep{kip90}.  The over-adiabacity, $x\equiv \nabla - \nabla_{ad}$,
of the stellar envelope depends on the mixing length parameter with a
range of powers from $x\propto l_{m}^{0.0}$ to $x\propto
l_{m}^{-1/3}$.  Thus, for two stars with the same central temperature,
the star with the smaller mixing length will always have a larger
temperature gradient leading to a cooler surface temperature and an
older LDB age.

\begin{figure}
\plotone{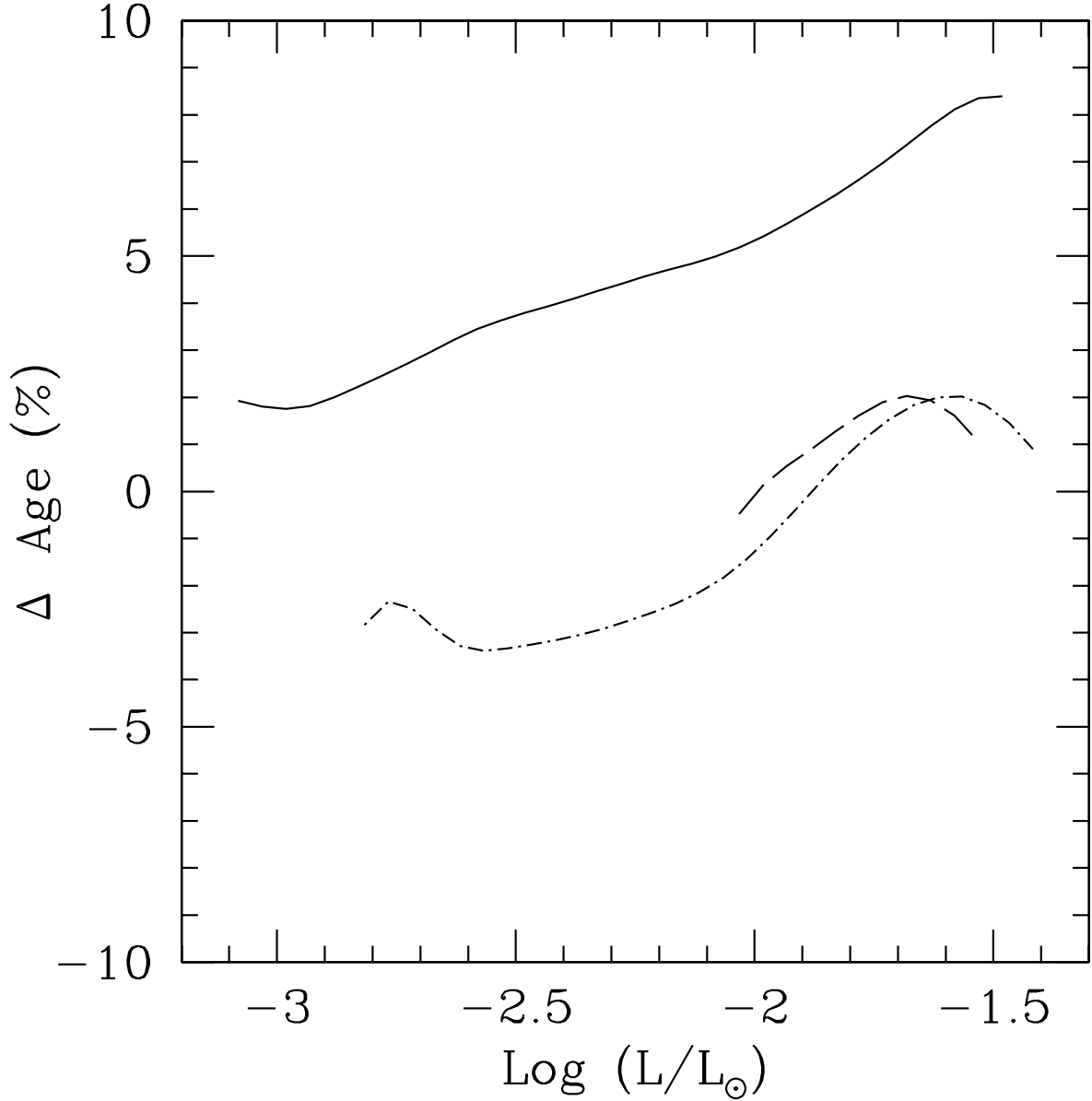}
\caption{Solid line shows the percentage difference in the LDB ages between adopting the mixing length parameter, $\alpha=1.0$ and the solar calibrated, $\alpha=1.75$, reference model.  A positive difference is in the sense that the parameter variation gives an older LDB age than the reference model.  Dashed line shows the percentage difference in the LDB ages obtained by adopting the MHD equation of state.  The upper density limit of the MHD equation of state restricts calculations to M/M$_{\odot}>0.175$.  The line has been shifted $\Delta Log(L)=-0.1$ for better visibility.  Dashed-dot line shows the percentage difference in the LDB ages obtained by adopting the OPAL 2001 equation of state \citep{rog02}.  The limits of the table restricts calculations to $M/M_{\odot}>0.08$.}
\label{mixlen}
\end{figure}

Figure~\ref{mixlen} quantifies the effect on the LDB age of reducing
the mixing length parameter from the reference model's value, $\alpha
= 1.75$, to $\alpha =1.0$.  The solid line in Figure~\ref{mixlen}
shows the difference in age between the reduced-mixing-length
LDB-luminosity-age relation and the reference LDB-luminosity-age
relation as a percentage of the reference model LDB age at fixed
luminosity.  All comparisons between models are at fixed luminosity,
since the luminosity of the LDB is the observed quantity.  A positive
difference is in the sense that reducing the mixing length parameter
leads to older LDB ages at fixed luminosity than the reference model.

The mixing-length formalism is a 1-D static approximation of a
phenomena that is intrinsically 3-D and time dependent.  Thus, it may
not be an accurate representation of convection.  Computational
limitations prevent a full 3-D parameter-free convection
implementation making it difficult to fully quantify the shortcomings
of the mixing-length formalism in determining the LDB ages.  Some
constraints can be placed by comparing our calculations to other
calculations in the literature that implement alternative 1-D static
convection treatments.  The LDB calculations of \citet{dan94}
implement the \citet{can91} convective theory and \citet{dan97}
implement the convection theory outlined in \citet{ven98}.  Our
calculation of enforcing the adiabatic temperature gradient
(Section~\ref{opac}), bypassing the convection theory, provides
another handle on the impact the convection theory has on LDB ages

For the remaining variations to the input physics, unless otherwise
stated, we recalibrate the mixing length and helium to match the solar
data.  The recalibration does not significantly affect the LDB
calculation.  For example, in the case of using a gray atmosphere (see
Section~\ref{atmos}), the LDB age difference at the highest luminosity
would be 4.5\% with a recalibration of the mixing length instead of
5.0\% without recalibrating the mixing length.

\subsection{Equation of State}\label{eos} The reference models employ
the \citet{igl96} equation of state and are supplemented toward lower
temperatures and higher densities by the \citet{sau95} equation of
state.  One alternative equation of state valid in the low mass regime
is the MHD equation of state by \citet{dap88}.  Unfortunately, the
density upper limit, $\log \rho = -2.0$, of the MHD equation of state
prevents calculations to the lithium depletion age for
$M/M_{\odot}<0.175$.  For consistency we recalibrate the mixing length
parameter, $\alpha =1.75$, and helium abundance, Y=0.272, to match the
solar data using the MHD equation of state.  The dashed line in
Figure~\ref{mixlen} shows the MHD equation of state results in older
LDB ages, $\sim 2\%$, over the restricted mass range.  This line is
offset slightly for clarity.

Recently, the OPAL group updated their equation of state.  The latest
equation of state, OPAL 2001, treats the electrons relativistically,
improves the activity expansion method for repulsive inter-particle
interactions, and covers a larger range in temperature and density
\citep{rog02}.  The dot-dashed line in Figure~\ref{mixlen} shows the
impact on LDB ages using the OPAL 2001 equation of state.  Despite the
improved coverage of temperature and density, the table is still
inadequate for evolution to the lithium depletion boundary for $M<0.08
M_{\odot}$.  As a side note, evolution to the main sequence is only
possible for $M>0.125 M_{\odot}$ using the OPAL 2001 equation of state
only.

Uncertainties in the low density, high temperature regime covered by
the \citet{igl96} calculations have a negligible impact on LDB ages.
Using the \citet{sau95} equation of state in place of \citet{igl96}
results in no change to the LDB age.  For completeness, we calculate
LDB ages using an older generation equation of state based on the Saha
equation with Debye-H\"{u}ckel corrections \citep{gue92}.  The Saha
equation of state results in 50\% older LDB ages over a limited mass
range, $M \ga 0.2 M_{\odot}$.  The calculation becomes numerically
unstable for lower masses.  Use of the Saha equation of state for a
solar model also results in a poor fit to solar oscillation data
\citep{gue92}.  For the above reasons, the Saha equation of state is
too simplistic for realistic calculations and is not included in the
overall error budget.

\subsection{Opacity}\label{opac} In stellar models, the radiative
opacity determines the temperature gradient in regions where energy is
transported by radiation and to a lesser degree determines the
temperature gradient in superadiabatic convective regions.  In the
mass range relevant to the LDB, efficient convection transports energy
over $\sim 98\%$ of the stellar mass.  Thus, radiative opacities
become important in determining the stellar structure of the remaining
2\% of the stellar mass.  Unfortunately, molecular dissociation and
ionization greatly complicate the opacity calculations in these outer
regions.

As shown in Figure~\ref{runrho}, the commonly used low temperature
opacities of \citet{ale94} are insufficient to model the
superadiabatic and possible radiative zones present in the low mass
stars of this study.  The diagonal dotted line shows the upper density
limit of the \citet{ale94} opacity data.  Thus, for the reference
models we supplement the \citet{ale94} low temperature opacity data
toward higher densities with the molecular opacities of \citet{ale83}
as described in Section~\ref{fiducial}.  Energy transport of the
stellar interior follows the adiabatic temperature gradient, so any
changes in the opacities at high temperatures, $\log$(T)$\ga 4.0$, are
inconsequential to the LDB.  Thus, only low temperature opacities are
varied in this study.  When varying the low temperature opacity of the
envelope, we are unable to use a self-consistent atmosphere that
employs the same alternative opacity.

\begin{figure}
\plotone{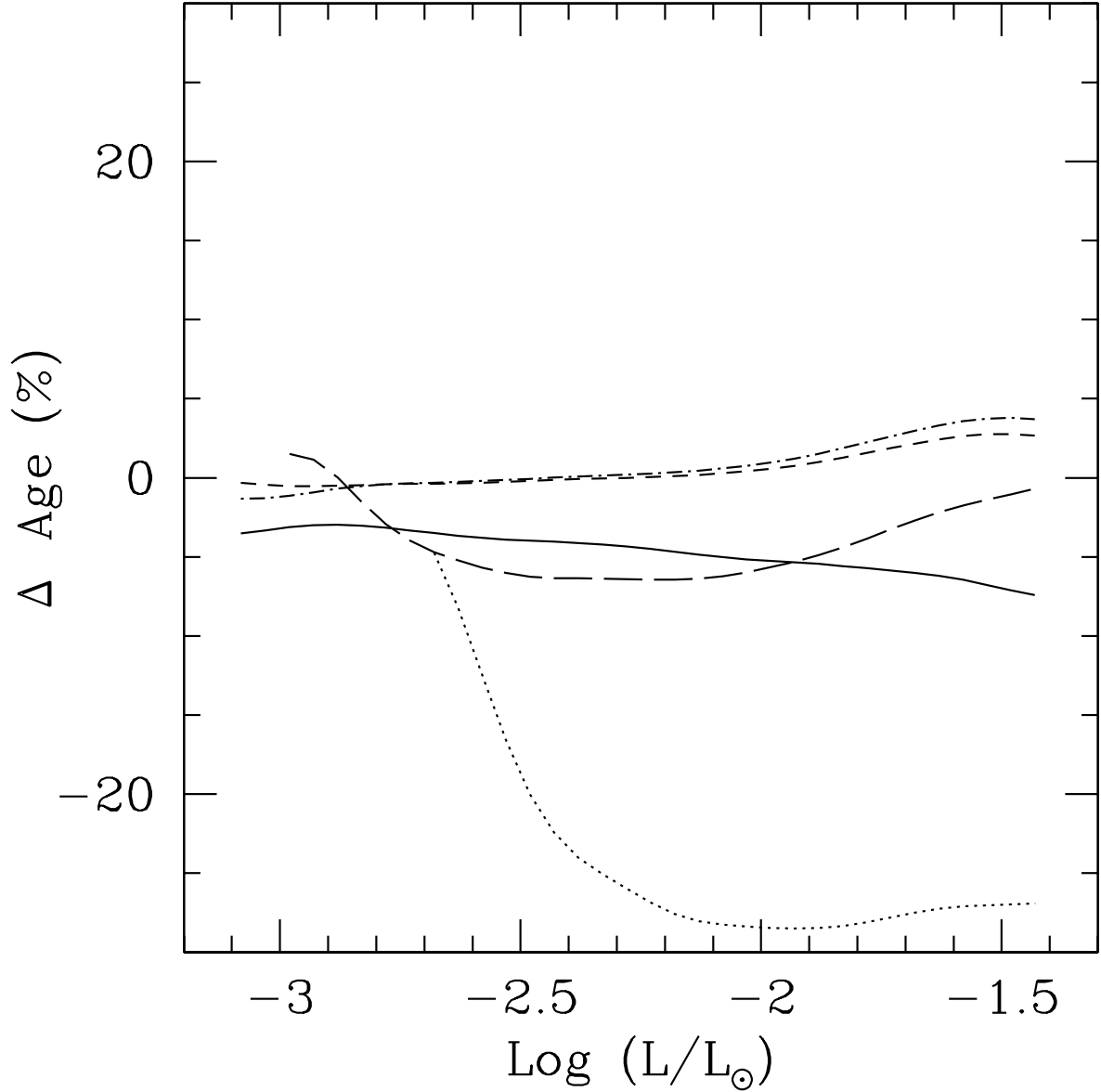}
\caption{Solid line shows the percentage difference in the LDB ages obtained by enforcing the adiabatic temperature gradient.  The dashed line shows the difference using the \citet{kur98} low temperature opacity.  The short-dashed line shows the LDB age difference using the \citet{all01} dusty opacities.  The dot-dashed line compares the LDB age using the high opacity.  The dotted line is for a zero metallicity opacity.}
\label{opacfig}
\end{figure}

As a limiting case, we show the effect of enforcing the adiabatic
temperature gradient throughout the stellar interior, essentially
ignoring the mixing length theory, as the solid line in
Figure~\ref{opacfig}.  The superadiabatic region is eliminated,
leading to a smaller temperature gradient and hotter star, which
evolves more rapidly.

For alternative low temperature opacities, we first considered
replacements to the more uncertain high density, $R>1$, molecular
opacities of \citet{ale83}.  However, the stellar envelope evolves
roughly perpendicular to the R=1 opacity boundary (see
Figure~\ref{runrho}).  Using opacity tables that contain large
systematic differences at the R=1 transition region can lead to rapid
variations in the entropy structure of the outer envelope.  The
increase or decrease in the size of radiative and superadiabatic zones
can enhance or reduce the rate of evolution as described in
\citet{gui95}.  Thus, patching two opacity tables together can
artificially affect the rate of evolution.  Fortunately, our selection
of tables in the reference models results in a smooth transition, and
the effect on our reference models is negligible.  To quantify the
impact, we calculate LDB ages with opacities toward higher densities
obtained by a linear-in-log-space extrapolation of the \citet{ale94}
opacities.  The difference in LDB ages is less than 1\% over the
entire mass range.

The above discussion emphasizes the importance of continued work in
the low temperature opacities for self-consistent calculations over a
wider range in density.  For the comparisons that follow, we use a
single opacity table and perform the linear extrapolation toward
higher densities if necessary.  The low temperature opacity table is
connected to the high temperature opacity at $\log$(T)=4.0, and the
high opacity and high density at the boundaries of these opacity
tables ensure convective transport of energy.  Additionally, hydrogen
ionization dominates the opacity at the boundary between the high and
low temperature opacities.  Thus, tables of differing metal abundances
match well.

\citet{kur98} provides an alternative opacity calculation in the low
temperature regime.  We show its impact on the LDB ages as the dashed
line in Figure~\ref{opacfig}.  The opacity data have the same R=1
boundary as the \citet{ale94} opacities, and are extrapolated toward
higher densities.  The mixing length and helium are recalibrated to
the solar data yielding $\alpha =1.731$ and $Y=0.272$, respectively.
The \citet{kur98} and \citet{ale94} opacities are very similar except
for $\log$(T)$<$3.6, where the effects of molecules become important.
The opacities of \citet{ale94} are increasingly larger for
$\log$(T)$<$3.6 in comparison to the \citet{kur98} opacity data.  The
lower \citet{kur98} opacities in the outermost layers lead to a hotter
star, which evolves more rapidly.

We compare our reference models to models using the latest opacity
calculations of \citet{all01} as the short-dashed line in
Figure~\ref{opacfig}.  Figure~\ref{opacfig} shows the comparison
between a calculation employing the so-called "dusty" opacities that
include the affects of dust grains on the opacities to the reference
calculation.  The "dusty" opacity table fully covers the low
temperature regime under investigation here.  The solar calibration
yields $\alpha =1.767$ and $Y=0.272$.  We also calculate LDB ages
using the so-called "condensed" opacities of \citet{all01} that remove
grains from the opacity since they condense out of the photosphere.
Since the models are never cool enough for grain formation, neither
the "dusty" opacities or "condensed" opacities appreciably affect the
LDB ages.

The opacity tables presented so far have qualitatively similar
structure.  The opacities have a maximum at $\log$(T)=4.4 due to
hydrogen ionization and sharply decrease toward cooler temperatures
with a minimum at $\log$(T)$\sim$3.3.  The opacities begin to increase
toward even cooler temperatures due to the formation of molecules and
collision-induced absorption.  Also, there is an overall increase in
the opacities toward higher density.  Following advances in more
complete line lists, the overall opacity tends to increase.  In order
to study the impact of higher opacities and the possibility of a
change in the qualitative shape to the opacity data, we create an
opacity table that is identical to the reference calculation opacities
for temperatures greater than $\log$(T)=4.0.  For temperatures less
than $\log$(T)=4.0, the opacities at fixed density are extrapolated
toward lower temperatures at a constant value given by the opacity
value at $\log$(T)=4.0 for the corresponding density.  Thus, the only
structure kept is the increase in opacity for increasing density.

The dot-dashed line in Figure~\ref{opacfig} shows the resulting LDB
ages using the extreme high opacity table described in the preceding
paragraph and a solar-calibrated $\alpha =2.096$ and $Y=0.271$.  The
higher opacities do increase the superadiabatic temperature gradient,
which results in cooler stars.  However, there is only a modest
increase in the LDB ages.  The opacities are currently high enough
that the stars are strongly convective all the way to the atmosphere
boundary point.

In the opposite opacity extreme, we calculate LDB ages for a
zero-metallicity opacity \citep{sta86}.  The solar calibration results
in $\alpha =1.50$ and $Y=0.271$.  The significant reduction in the LDB
ages for the zero-metallicity opacity is shown as the dotted line in
Figure~\ref{opacfig}.  The effective temperature is 20 to 30\% larger
than the reference case.  These low and high opacities are obviously
extreme and unphysical, but they set limiting cases.

One additional note on the topic of opacity, we confirm the findings
of \citet{cha00}, that conductive energy transport only affects later
stages of evolution than the lithium depletion and has no impact on
the LDB ages.

\subsection{Atmosphere \& Boundary Condition}\label{atmos} Even though a standard gray
atmosphere results in unrealistically high effective temperatures in
the mass regime studied here, it provides an extreme case for
deviations from the reference calculation, which employs the
\citet{hau99} atmosphere.  For the gray atmosphere, we use the
reference calculation's mixing length and helium.  The impact on the
lithium depletion relation is shown as the solid line in
Figure~\ref{atmosfig}.  The LDB age is 5\% younger for the highest
luminosity and is negligibly different for the lowest luminosity.
There are few alternatives to the gray atmosphere and \citet{hau99}
atmosphere for modeling low-mass stars.  Comparisons to other
independent LDB calculations in Section~\ref{exterr} provide some
constraints on the impact alternative atmospheres can have on the LDB
ages.

The \citet{hau99} atmosphere limits calculations to solar metallicity.
To check the impact of metallicity variations on the LDB ages, we use
the gray atmosphere for a nonsolar-metallicity calculation.  A
gray-atmosphere calculation using a metallicity that is increased by
0.1 dex (assuming $\Delta$Y/$\Delta$Z=2.0) results in a less than 2\%
increase in the LDB ages over all luminosities when compared to the
gray-atmosphere calculation at solar metallicity.

The atmosphere provides the outer boundary conditions required for the
stellar interior calculation.  Our reference calculation fixes the
boundary condition at an optical depth where the temperature is equal
to the effective temperature.  This optical depth is close to
$\tau=2/3$.  In their calculation of low mass stellar models,
\citet{bar98} choose a deeper $\tau=100$ boundary condition.  They
prefer the deeper boundary condition to ensure the interior
calculation is in the fully-adiabatic regime and to reduce the
extrapolation of the opacity data.

It is unclear which of these fitting points is more appropriate.  In
our case of the shallow $\tau=2/3$ boundary condition, the potential
of a mismatch between the interior and atmosphere opacity data
resulting in abrupt variations to the convective stability criterion
or overadiabicity exists.  Whereas the deeper $\tau=100$ boundary
condition reduces this problem, it has a disadvantage since the
atmosphere calculations of \citet{hau99} do not include real gas
effects in the equation of state.  The mismatch between the interior
\citet{sau95} equation of state and \citet{hau99} equation of state
can be as large as 1.6\% in density at $\tau=100$, whereas the
mismatch is $<$0.9\% in density at $\tau=2/3$.  In summary, both of
these fitting points are equally valid and neither is ideal.  To
quantify the impact on the LDB ages due to variations in the boundary
condition depth, we calculate the LDB using $\tau=100$.  This results
in a 3\% younger LDB for the oldest clusters and a 10\% older LDB age
for the youngest clusters (dashed line - Figure~\ref{atmosfig}).

\subsection{Rotation}\label{rotsec} Rotation affects LDB ages in several ways.  The change in kinetic energy of rotation enters the energy
generation equation, and rotation provides additional pressure support
which can alter the effective temperature and thus the rate of
gravitational contraction.  Current observations and theory suggest
the rotational evolution of a star begins with efficient angular
momentum loss via a disc-locking mechanism to nearly constant angular
momentum evolution to again efficient angular momentum loss via a
stellar wind \citep{sta03}.  Due to the complicated rotational
evolution of a star we investigate two limiting cases of evolution at
constant angular momentum and evolution with significant amounts of
angular momentum loss.  As a guide for these two cases we match the
model rotation rates to the observed upper envelope of rotation rates
in the Pleiades open cluster \citep{ter00}.  The upper envelope of
rotation rates is approximated by a linear decrease in $V\sin i$ from
$V\sin i$=70 kms$^{-1}$ at M=0.3 M$_{\odot}$ to $V\sin i$=50
kms$^{-1}$ at M=0.1 M$_{\odot}$ (see Figure 7 of \citet{ter00}).  The
model rotation rates are extrapolated toward lower masses.  We assume
$\sin i=\pi /4$ to obtain the physical rotation rate.

For evolution without angular momentum loss, rotation rates are
matched to the Pleiades rotation rates at an age of 120 Myr.  Since
the stars evolve at constant angular momentum and spin up as their
radii contract, adopting an older age for the Pleiades
(self-consistent with the LDB age derived in Section~{conclude}) would
reduce the impact of rotation even further.  The no angular momentum
loss evolution results in 2\% older ages for the minimum LDB mass and
$<$ 1\% older ages for higher masses.  Previous claims of 20\% older
LDB ages result from allowing a rotation rate greater than 200
kms$^{-1}$ at the Pleiades age for the minimum LDB mass \citep{bur00}.
Such high rotation rates are only observed for solar-mass stars, and
our adopted 65 kms$^{-1}$ rotation rate at Pleiades age in this study
is more appropriate for the lower mass stars relevant to the LDB ages.

Our second limiting case for the rotational evolution examines the
impact of efficient angular momentum loss.  The stars are forced to
evolve at a constant rotational velocity matched to the Pleiades
rotation rates as described in the preceding paragraph up to an age of
120 Myr and henceforth evolve at constant angular momentum.  This
results in 2\% younger ages for the minimum LDB mass and 6\% younger
ages for the highest LDB mass.

Open cluster members of a given mass have a wide range of rotation
rates which indicates a variety of evolutionary sequences
\citep{ter00}.  Thus an open cluster of a given age may have a less
well defined LDB as a result of stars with differing rotation rates
depleting lithium slightly earlier or later than our reference case.
Our limiting cases for the angular momentum evolution allows us to
quantify this rotational ``smearing'' of the LDB location.  The
rotational LDB ``smearing'' is quantified by calculating the
difference in bolometric magnitude of the lithium depletion at fixed
age.  By comparing our no angular momentum loss calculation to the
efficient angular momentum loss calculation, we find the difference in
bolometric magnitude at fixed age between these calculations is 0.12
and 0.04 mag for young and old clusters, respectively.

In conclusion, since these calculations that include rotation are
limiting cases these deviations are conservatively considered as
2-$\sigma$ effects.  Thus, rotation negligibly affects the LDB
calculation and extent of the observed LDB in comparison to the other
sources of theoretical and observational errors.  Thus, we do not
include a contribution of these effects in the overall error budget
for the LDB age technique.

\section{Internal Error Budget}\label{interrbudget} Input physics that
affect the superadiabatic and radiative regions of the atmosphere
dominate the uncertainty in the LDB ages.  Qualitatively, results from
the previous section imply that the uncertainty in the lithium age
dating technique is larger for the higher masses, even though the
input physics for the higher masses are relatively more secure.  This
behavior in the LDB uncertainty is explained in the following way.
For a given set of physical inputs, the LDB age as a function of
luminosity is a sequence of mass.  However, for the lowest masses,
variations in the effective temperature of a star move the resulting
LDB age--and luminosity--parallel to the sequence in mass.  This
effect results in the apparent robustness of the LDB ages to
variations in the effective temperature at the low luminosity, low
mass end.

\begin{figure}
\plotone{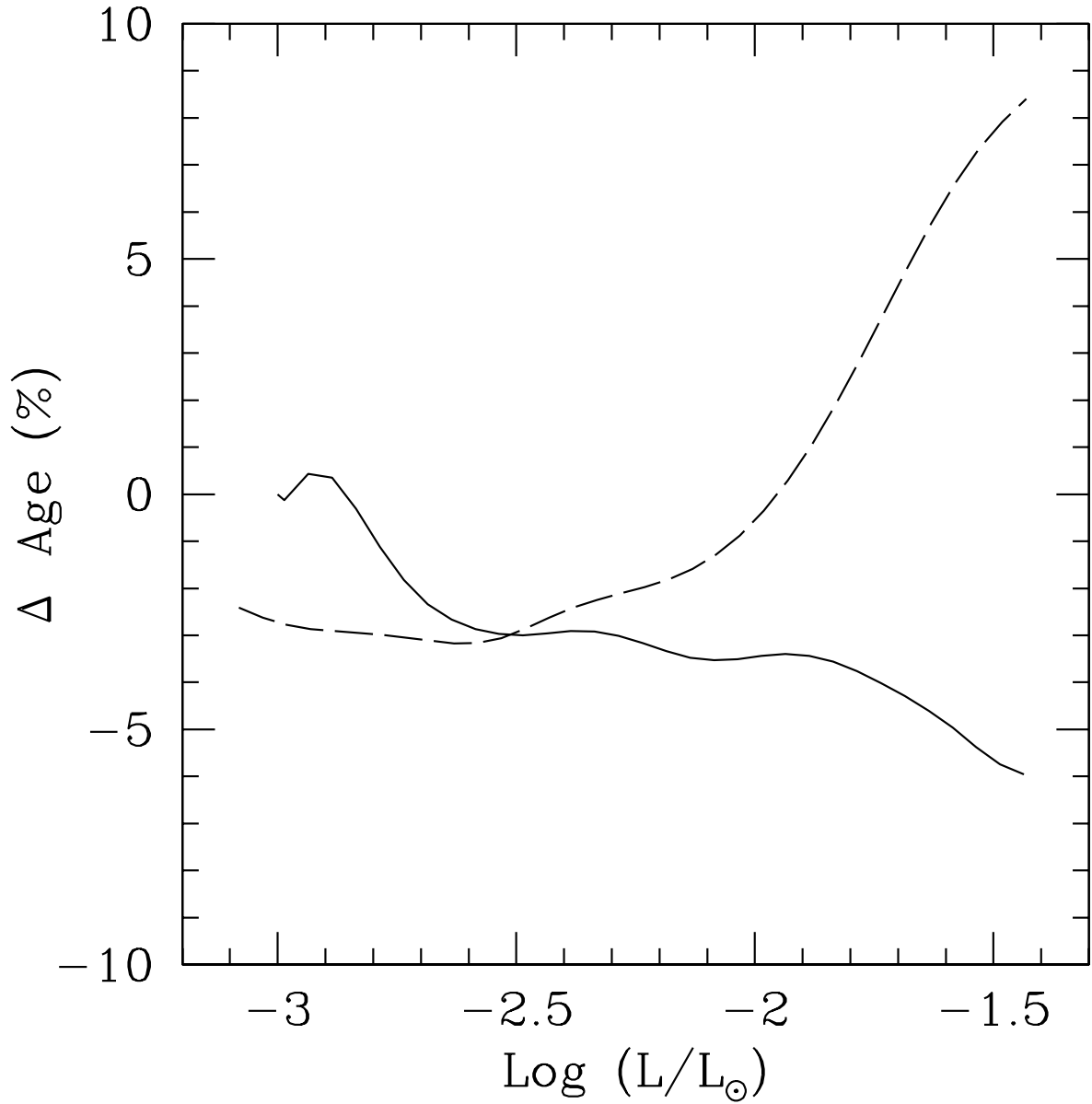}
\caption{Solid line shows the percentage difference in the LDB ages obtained by using the gray atmosphere.  Shows the LDB age adopting a $\tau=100$ fitting point for the boundary condition.}
\label{atmosfig}
\end{figure}

The robustness of the LDB ages is best illustrated for the gray
atmosphere case (solid line - Figure~\ref{atmosfig}).  Using a gray
atmosphere, the $M=0.065\, M_{\odot}$ model has a 7\% higher effective
temperature than the reference model.  The 7\% increase in the
effective temperature results in a 15\% decrease in the time for
lithium depletion and a 24\% increase in the luminosity at the time of
lithium depletion.  The changes in the lithium depletion luminosity
and age are almost identical to changes resulting from a variation in
mass.  The gray atmosphere $M=0.065$ M$_{\odot}$ star has identical
lithium depletion luminosity and age as a $M=0.068$ M$_{\odot}$ star
using the reference \citet{hau99} atmosphere.  Thus, at fixed
luminosity, there is no age difference between these drastically
different atmosphere calculations for the lowest mass stars.

Combining the above results to arrive at a theoretical uncertainty in
the lithium age-dating technique is difficult.  We have treated all
variations in the input physics as independent and have not explored
correlations between the various input physics or the possibility of
nonlinear behavior.  Additionally, variations in mixing length,
opacity, and atmosphere can all be viewed as variations in the
effective temperature.  The effective temperature, second only to
mass, determines the rate of collapse.  Keeping the above in mind, we
first attempt to quantify the errors in the LDB ages using the results
from Section~\ref{interr}, where we quantify differences in the LDB
ages from our reference model that result from changes to the input
physics.

The impact on the LDB ages resulting from changes to the input
physics, as shown in Figures~\ref{mixlen}-~\ref{atmosfig}, provides
the basis for characterizing the error in the LDB ages as a function
of $\log$(L).  Because of the extreme deviation from a realistic
change to the input physics, we reduce the deviation for the case of
using a zero-metallicity, low temperature opacity (dotted line -
Figure~\ref{opacfig}) by 80\%, since a 20\% metallicity error is more
likely.  Assuming the deviation scales linearly with metallicity is
conservative since we find a 0.1 dex metallicity variation results in
less than 2\% age differences overall (see Section~\ref{atmos}).
Among the variations to the input physics, there is no convincing
evidence for a preference between older or younger LDB ages in
relation to our reference model.  Thus, we adopt a symmetric error.
We adopt the absolute value of the deviations shown in
Figures~\ref{mixlen}-~\ref{atmosfig} as 2-$\sigma$ errors.  For
individual error sources that do not extend the full luminosity range,
-3.0$\la \log$(L/L$_{odot}$)$\la -1.5$, we linearly extrapolate the
errors to a value of zero at both luminosity extrema.  By adding these
deviations in quadrature, we arrive at the total 1-$\sigma$ error in
the LDB ages as a function of $\log$(L/L$_{\odot}$), shown as the
solid line in Figure~\ref{errfig}.

\section{Independent Lithium Depletion Boundary
Comparisons}\label{exterr} Another way to estimate the systematics in
the LDB ages is to compare our result with other independent
calculations from the literature.  As a first comparison, we compare
the reference calculation to the \citet{bil97} analytical model for
lithium depletion.  The analytical treatment by \citet{bil97} requires
several assumptions.  They assume that the temperature gradient follows
the adiabatic temperature gradient for a fully ionized gas,
$\nabla_{ad}=2/5$.  Since the largest uncertainty in low mass stellar
models arises from determining the effective temperature,
\citet{bil97} treat the effective temperature as a free parameter that
is constant during pre-main-sequence contraction.  To integrate the
lithium destruction over the stellar interior, \citet{bil97} expand
the stellar structure around the central value, and they ignore the
change in degeneracy and electron screening as a function of the
interior entropy.

\begin{figure}
\plotone{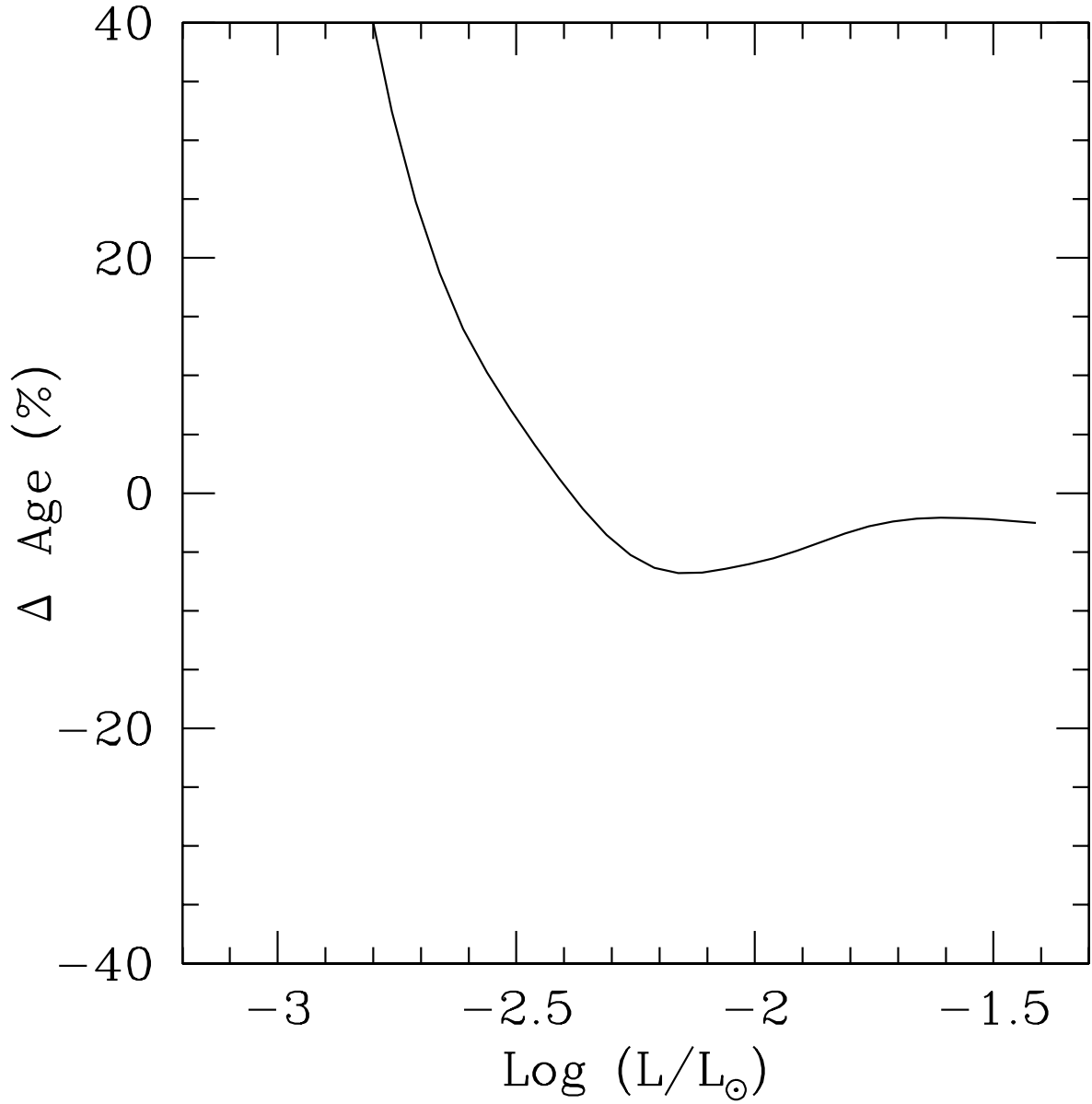}
\caption{Percentage difference in the LDB ages between the analytical model of \citet{bil97} and the numerical reference model.  A positive difference is in the sense that the analytical model gives an older LDB age than the reference model.}
\label{analyt}
\end{figure}

Figure~\ref{analyt} shows the percentage age difference at fixed
luminosity between the \citet{bil97} analytical model (their eq. 11)
and our reference calculation for the LDB-age relation.  A positive
difference is in the sense of the analytical model having older ages
at fixed luminosity than the reference calculation.  To calculate the
analytical-model LDB-age relation, we use the same effective
temperature and effective molecular weight, $\mu_{eff}\equiv\rho N_{A}
k_{B} T/P$, at lithium depletion as obtained from the reference
calculation and assume a 99\% lithium depletion level.  The
analytical-model results are limited to $M>0.08 M_{\odot}$.  At the
high luminosity end, the analytical model underestimates the LDB age
by $\sim$ 5\%.  For $M\la 0.1 M_{\odot}$ ($\log (L/L_{\odot})\la -2.2$), the
effects of degeneracy become important, and the analytical model
begins to overestimate the age.  Thus, the analytical model follows
the numerical results within a numerical constant when degeneracy is
not important, but to extend the LDB age technique beyond 90 Myr
requires a fully numerical calculation.

\begin{figure}
\plotone{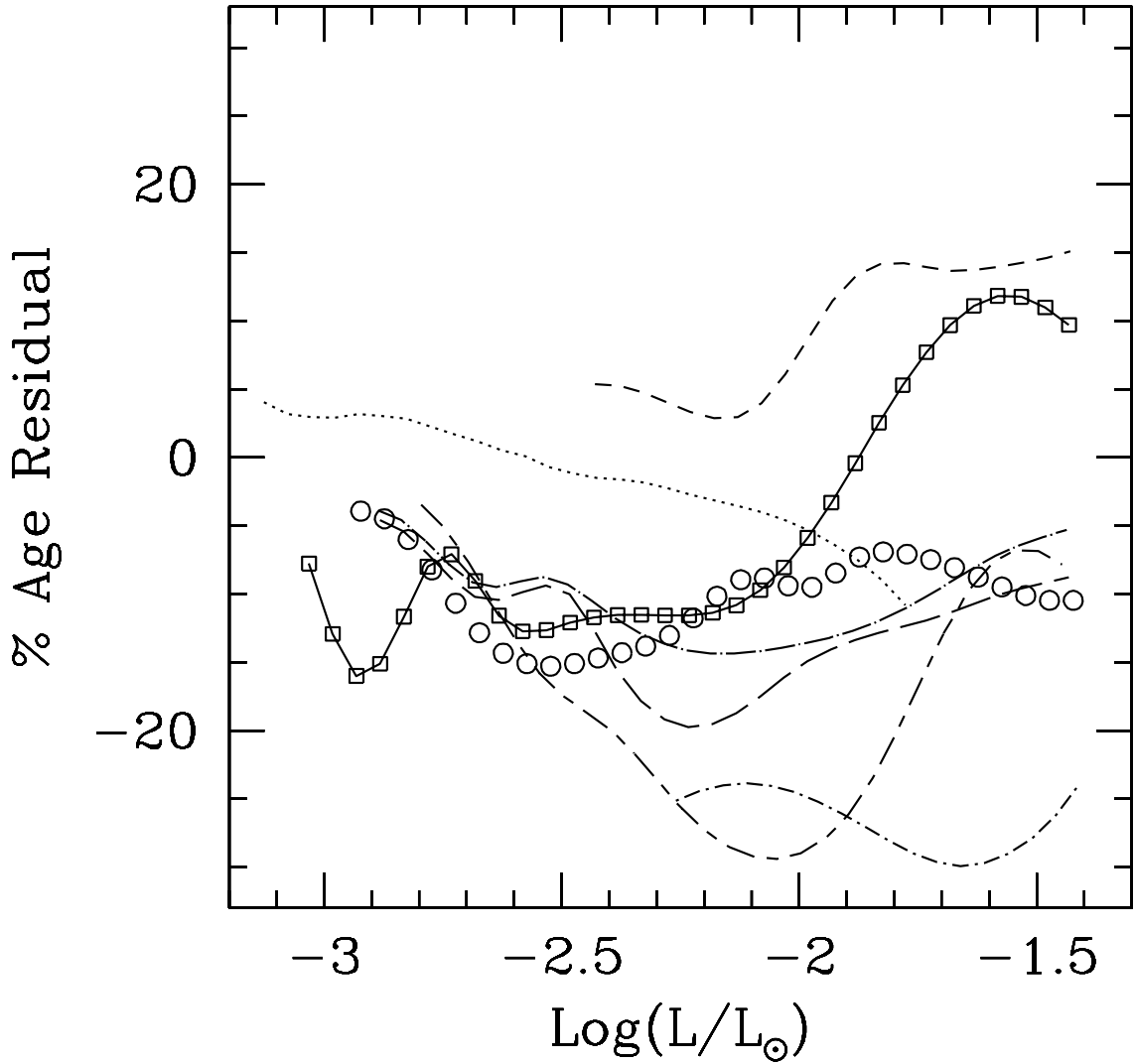}
\caption{Percentage difference in the
LDB ages between the reference model and independent LDB age
calculations from the literature.  Short-dashed line is the comparison
with \citet{sie00}.  Dotted line is the comparison with \citet{bur97}.
The four remaining lines are comparisons to different model
assumptions in \citet{dan94}.  The solid line with open squares
delineate the comparison with the calculations of \citet{bar98}, and
the open circles shows the comparison to the calculation of
\citet{dan98}.}
\label{comp}
\end{figure}

The comparison with the analytical model demonstrates the robust
nature of the LDB age technique.  For an additional characterization of
the error in LDB ages, we only use full numerical calculations from
the literature that treat the relevant physics more accurately.
Figure~\ref{comp} shows the difference in the lithium depletion age at
fixed luminosity for several other calculations.  The dotted line
compares our results to the calculations of \citet{bur97}, the
short-dash line is the comparison to \citet{sie00}, and the four sets
of lines with negative age residuals are comparisons to models in
given \citet{dan94}.  The solid line with open squares delineates the
comparison with \citet{bar98}, and the open circles show the
comparison with the ''1998 UPDATE'' models of \citet{dan98}.  With the
exception of the calculations from \citet{bar98}, the variations are
similar to the trend seen in the internal error budget; the residuals
are small at the faint end and increase toward higher masses.

Of the comparisons from the literature, the input physics of
\citet{bar98} are the most similar to ours.  The input physics of
\citet{bar98} differ from our reference calculation by their use of
the \citet{sau95} equation of state only, adoption of a mixing length
parameter, $\alpha=1.0$, and a deeper $\tau$=100 fitting point of the
outer boundary condition.  We do not find any impact on the LDB ages
as a result of using only the \citet{sau95} equation of state (see
Section~\ref{eos}).  Our adoption of a mixing length parameter,
$\alpha=1.0$, results in older LDB ages (see Section~\ref{mixlen}).
Our adoption of the $\tau=100$ fitting point results in older LDB ages
for the youngest clusters, and slightly younger LDB ages for older
clusters (see Section~\ref{atmos}).  The calculations of \citet{bar98}
deviate from our reference calculation in a manner that qualitatively
resembles our calculation adopting the $\tau=100$ fitting point.
However, the LDB ages of \citet{bar98} at $\log$(L/L$_{\odot}$)=-2.9
are 5-$\sigma$ younger than our expectations based on the internal
error-budget calculation (Figure~\ref{errfig}).

The input physics of the \citet{dan94} models differ from our
reference model in their use of the MHD equation of state
\citep{dap88} for densities, $\rho <0.01$ gcm$^{-3}$ and the
\citet{mag79} equation of state otherwise, and a gray atmosphere.  The
four sets of models compared in Figure~\ref{comp} are combinations of
using either the \citet{ale94} or \citet{kur98} low temperature
opacities and either the mixing length convection theory or the
\citet{can91} convection theory.  Overall, the four models result in
younger LDB ages.  The long-dash-dot line shows the comparison that is
closest to our reference calculation input physics (\citet{ale94}
opacity and mixing length convection theory).  Comparing this model to
our gray atmosphere calculation (solid line - Figure~\ref{atmosfig})
reveals a remaining difference of 10\%.  Ignoring implementation
differences, we infer the impact of using the \citet{mag79} equation
of state is at most 10\%.  The long-dash-small-dash line shows the
comparison where the only change with the previous model is the
\citet{kur98} low temperature opacity.  The largest age difference
occurs at $\log$(L/L$_{\odot}$)=-2.1, and this mimics our result when
using the \citet{kur98} opacity (dashed line - Figure~\ref{opacfig}).
The long-dash line is the comparison for the \citet{dan94} model that
uses the \citet{ale94} opacity and \citet{can91} convection theory.
The short-dash-dot line shows the difference when using the
\citet{kur98} opacity and \citet{can91} convection theory.  The
published data for the previous set of input parameters end at $M=0.1
M_{\odot}$.  The only change to the input physics relevant to the LDB
age technique between \citet{dan97} and \citet{dan94} is the
convection treatment \citep{ven98}.  As we see in the comparison to
\citet{dan97} (open circles), the alternative convection treatment has
little impact on the LDB ages.

The dotted line compares our reference calculation to that of \citet{bur97}.
They employ their own opacity and atmosphere calculations but still have
very similar results to our reference calculation. The short-dash line
compares our reference calculation to that of \citet{sie00}.  For the
atmosphere, they use an analytical fit to the $T(\tau)$ relation
calculated by \citet{ple92}.  Their equation of state is based on the
framework as outlined in \citet{pol95}.

\section{External Error Budget}\label{exterrbudget} An alternative
method to calculate the LDB age uncertainty is to use external
comparisons with independent lithium depletion calculations from the
literature (see Figure~\ref{comp} and Section~\ref{exterr}).  We adopt
the absolute value of the deviations shown in Figure~\ref{comp} as
1-$\sigma$ errors.  We average the individual deviations to arrive at
the total 1-$\sigma$ error shown as the dashed line in
Figure~\ref{errfig}.  The four calculations from \citet{dan94} are not
included in the average since they have been superseded by the more
recent \citet{dan97} calculations.  For comparisons that do not extend
over the entire range of luminosity, we extrapolate by adopting a
constant value equal to the comparison endpoints.

The external 1-$\sigma$ error is larger than the internal 1-$\sigma$
error (Figure~\ref{errfig}.  The larger external error suggests there
are additional systematic deviations in the LDB calculations that
result from either numerical differences between the LDB calculations
or our calculations have not fully explored the range of relevant
physics.  However, reconciling the two error estimates only requires
an additional 5\% error added in quadrature to the internal error
estimate.  Since we are interested in the impact of input physics
uncertainties on the LDB ages only, for the 1-$\sigma$ theoretical
error in the LDB ages, we adopt the internal 1-$\sigma$ errors.
Adopting the larger external error may include numerical effects into
our error budget.  For a simple analytical model of the 1-$\sigma$
theoretical error, we adopt a linear trend in $\log$(L) starting at a
value of 8\% for $\log$(L/L$_{\odot}$)=-1.5 and decreasing to 3\% for
$\log$(L/L$_{\odot}$)=-3.0.  For luminosities beyond the above
luminosity range, we adopt a constant percentage error equal to the
endpoint values.  The column labeled $\sigma_{THY}$ in
Table~\ref{reftab} gives the 1-$\sigma$ error in the reference LDB age
as a function of luminosity based on our simple analytical error
model.

Taking into account theoretical errors alone, absolute ages of open
clusters accurate to 3\% for the oldest clusters increasing to 8\% for
the youngest clusters using the LDB technique are possible.  Our
calculations of the errors in the LDB age technique show this to be
the most accurate method for determining the ages of open clusters.
These small errors emphasize the simplicity for calculating the
evolutionary rate of fully convective objects powered only by
gravitational energy.  Even LDB calculations that ignore essential
physics (analytical model of \citet{bil97}, ideal gas equation of
state, and zero metallicity opacities) have 1-$\sigma$ deviations from
our reference calculation that are of order 12\%.  Thus, unless we are
ignoring input physics that impact stellar structure of order the
error associated with treating the gas physics as an ideal gas, our
errors are accurate and are not underestimates.

\section{Bolometric Correction}\label{bolosec} Transformation from a
theoretical luminosity or effective temperature to an observable
magnitude system requires accurate knowledge of the bolometric
correction.  A quantitative discussion on the impact of the bolometric
correction uncertainty on the derived LDB ages does not exist in the
literature.  Thus, in this section we briefly characterize this
additional error source.  There are three methods for bolometric
correction determination: empirical, theoretical, and mixed
(theoretical method with empirical constraints).  Observations for
stars of known distance and wide spectral coverage allow an empirical
relation between the bolometric magnitude of a star and its broad-band
colors \citep{bes91,mon92}.  With improved atmospheres, fully
theoretical calculations for the bolometric corrections are possible
\citep{hau99}.  A mixture of the theoretical and empirical
methods attempts to improve the coverage of bolometric corrections
where observations are sparse, by collecting theoretical bolometric
calculations and then correcting them empirically \citep{lej97}.

For our reference bolometric corrections, we choose the theoretical
bolometric corrections of \citet{hau99}.  At fixed effective
temperature, the top panel in Figure~\ref{bcerr} shows the comparison
between our reference bolometric corrections and four other bolometric
correction determinations.  The solid line represents the comparison
with the mixed theory/empirical calculations of \citet{lej97}.  The
difference is in the sense of \citet{lej97} minus the bolometric
corrections of \citet{hau99}.  The long-dash line shows the comparison
with the theoretical models used by \citet{lej97} before applying the
empirical corrections.  The dash-dot line and short-dash lines compare
the results to empirical relations of \citet{mon92} and \citet{bes91},
respectively.  The empirical corrections, in practice, are given as a
function of observed photometric color, and blackbody spectral energy
distributions fit to the stellar spectra provide the effective
temperature scale.  Due to the departure from a blackbody, the
blackbody fitting method for determining the effective temperature
becomes increasingly inaccurate and highly questionable for
temperatures $<$ 3000 K \citep{mon92}.

The effect on LDB ages for the choice of bolometric correction is
shown in the bottom panel of Figure~\ref{bcerr}.  The bolometric
correction differences as a function of effective temperature are
mapped to LDB luminosity using the effective temperatures at lithium
depletion given in Table~\ref{reftab}.  For the 1-$\sigma$ error in
LDB ages due to bolometric corrections, we take the average absolute
value of the percentage age differences shown in the bottom panel of
Figure~\ref{bcerr} as 1-$\sigma$ deviations.  Unlike for the input
physics variations, it is not clear which bolometric correction is the
best choice.  Thus, the bolometric correction error budget is more
conservative than the theoretical error budget.

Additionally, there are two sources of systematic error in the
bolometric correction when applied to the LDB age technique.  Lithium
depletion occurs during the pre-main-sequence contraction phase.
Thus, the stars at the LDB have smaller $\log$(g) values than main
sequence stars.  Since empirical bolometric corrections are derived
from main sequence stars, the bolometric corrections are
systematically different.  Theoretical models used by \citet{lej97}
show the bolometric corrections roughly vary by 0.05 magnitudes per
0.5 dex change in $\log$(g).  The reference calculations have
$\sim$0.4 dex larger $\log$(g) values after 1 Gyr.  Using bolometric
corrections of $\log$(g)=5.0 for stars that actually have
$\log$(g)=4.5 results in a 2\% overestimate of the LDB ages across all
luminosities.

The second source of systematic error in the bolometric correction
results from the potential for the stellar cluster to differ in
metallicity from the empirical stellar templates employed in
determining the bolometric correction.  Bolometric corrections vary as
a function of metallicity especially for temperatures $<$ 3000 K.  The
theoretical models used by \citet{lej97} predict a 0.1 dex change in
[Fe/H] results in a 4\% difference in the resulting LDB ages for the
lowest luminosities.  The effect is less than 2\% toward higher
luminosities.  These two sources of systematic uncertainty are added
in quadrature with the dominant error source based on intercomparisons
between alternative bolometric correction determinations as discussed
in the preceding paragraph.  The resulting 1-$\sigma$ error is shown
as the long-dash line in Figure~\ref{errfig}.

A comparison of theoretical isochrone fits to the observed
color-magnitude diagram of the Pleiades is an additional check to the
systematic errors that may be present in the I-band bolometric
corrections.  For the coolest observed Pleiades members,
Teff$\sim$3400 K, the isochrone fit is too blue by 0.30 magnitudes in
(V-I) when using the \citet{hau99} bolometric corrections.  The
details of this isochrone fit to the Pleiades using the YREC stellar
evolution code is given in \citet{pin98}.  Similar sized discrepancies
are generic to all isochrone fits to open cluster color-magnitude
diagrams \citep{gro03}.  If the I-band bolometric correction is solely
responsible for the discrepancy in the color-magnitude diagram,
revisions to the bolometric corrections of a similar order could be
expected.  The correction is in the sense of reducing the LDB ages by
19-13\%, with the larger impact occuring at the low luminosity/old age
end.  Alternatively, the (V-I) colors of the isochrones can be
reconciled with the observations by a reduction of Teff$\sim$150 K in
the isochrones themselves.  Reducing the effective temperatures of the
reference calculation by 150 K results in a 12\% age reduction for the
oldest clusters decreasing to a 4\% age reduction for the youngest
clusters.  The reduction in LDB ages is for the bolometric correction
only and is independent of the effect the cooler temperatures would
have on the theoretical LDB ages.

For a simple analytical 1-$\sigma$ error in the LDB ages resulting
from uncertainties in the bolometric correction, we adopt a constant
error of 6.0\% for $\log$(L/L$_{\odot}$)$>$-2.6 and increasing
linearly to 10\% at $\log$(L/L$_{\odot}$)=-3.0.  The column labeled
$\sigma_{BC}$ in Table~\ref{reftab} gives the 1-$\sigma$ error in the
reference LDB ages as a function of luminosity.  The last column in
Table~\ref{reftab} gives the absolute I-band magnitude of the LDB
using the bolometric corrections of \citet{hau99}.  We adopt
$M_{bol\, \odot}$=4.74 and reduce the bolometric correction by 0.04
magnitudes to account for the systematically lower gravity at lithium
depletion.

\section{Conclusion}\label{conclude} 
\begin{figure}
\plotone{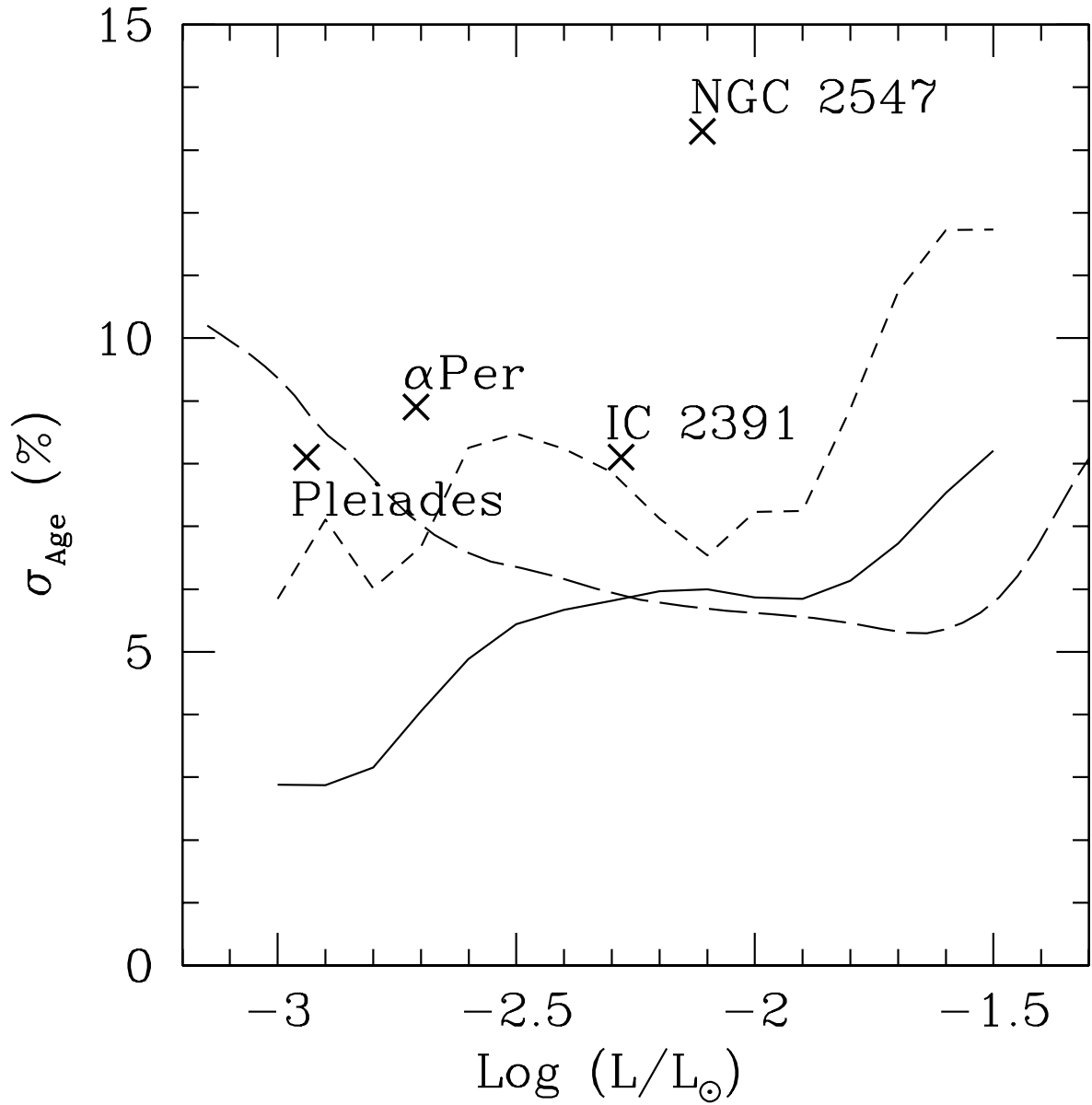}
\caption{Percentage, 1-$\sigma$ uncertainty in the LDB age as a function of $\log$(L/L$_{\odot}$).  Solid line and short-dash line shows the theoretical uncertainty based on deviations from the reference model for changes to the input physics and for comparisons to independent calculations from the literature, respectively.  Long-dash line is the uncertainty that results from bolometric correction uncertainties.  Crosses point out the observed luminosity of the LDB and the age uncertainty resulting from the uncertainty in its location for the labeled open clusters.}
\label{errfig}
\end{figure}

\begin{figure}
\plotone{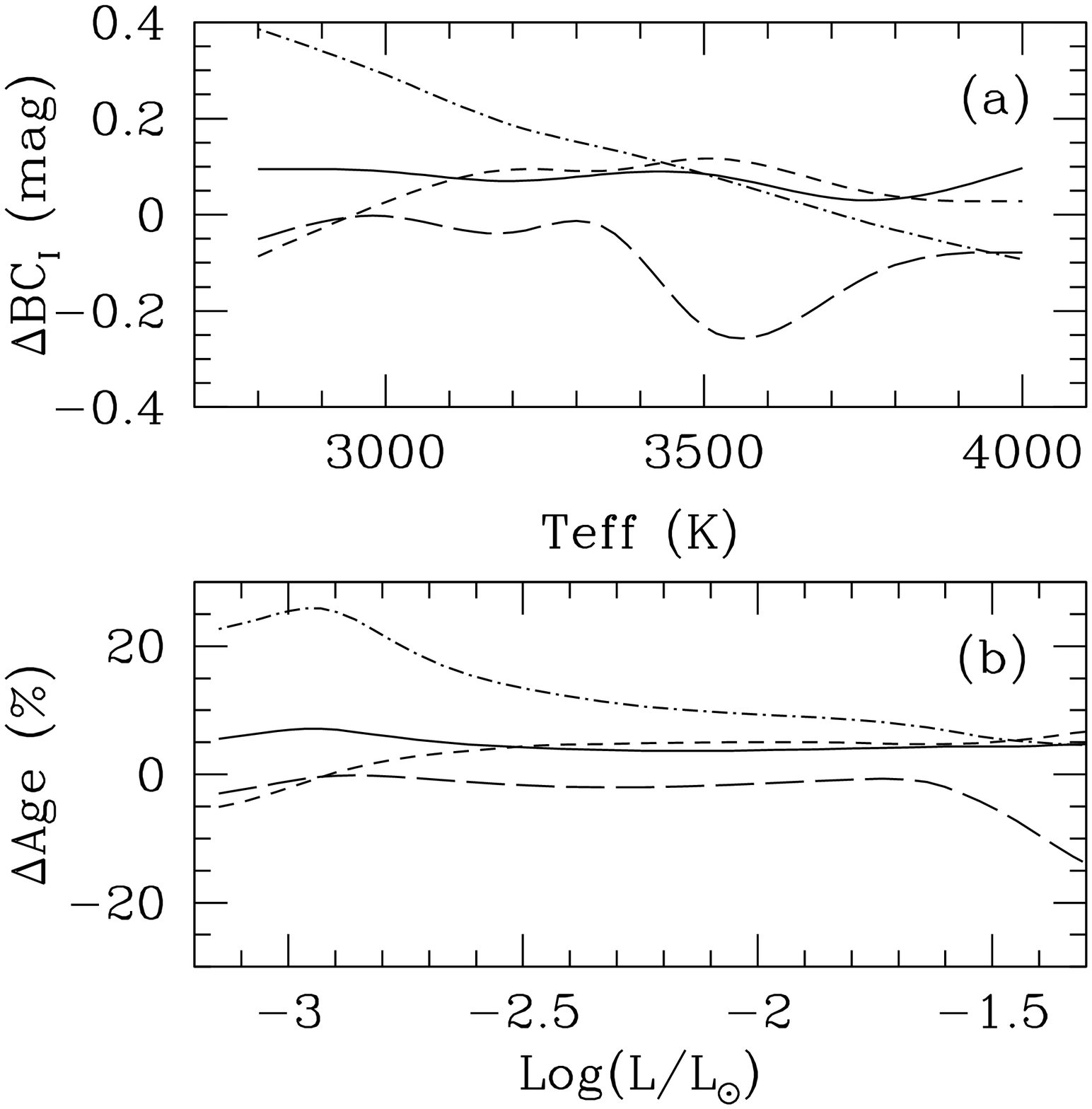}
\caption{ {\bf a} Magnitude difference in the I-band bolometric
correction at fixed effective temperature between several other
determinations and the theoretical calculations of \citet{hau99}.  The
solid line compares with the empirically corrected calculations of
\citet{lej97}.  The difference is in the sense of \citet{lej97} minus
\citet{hau99}.  The long-dash line shows the comparison with the
theoretical models used by \citet{lej97} before applying the
empirical corrections.  The dash-dot line and short-dash lines compare
the results to empirical relations of \citet{mon92} and \citet{bes91},
respectively.  {\bf b} Percentage difference in the LDB
ages at fixed $\log$(L/L$_{\odot}$) in comparison to the adopted
\citet{hau99} bolometric corrections.  Line types are identical to the
top panel.}
\label{bcerr}
\end{figure}

In conclusion, we employ our
error models for the theoretical and bolometric correction
uncertainties to rederive LDB ages and errors for the four clusters
with previous LDB age determinations (Pleiades - \citet{sta98};
$\alpha$ Per - \citet{sta99}; IC 2391 - \citet{bar99}; NGC 2597 -
\citet{oli03}).  We adopt the same apparent I magnitude of the LDB,
distance modulus, and reddening as in the original study.
Table~\ref{obstab} repeats the observational parameters, along with
their adopted errors.  We combine the observational errors of the LDB
location, reddening, and absolute distance modulus in quadrature to
derive the error in the absolute I magnitude of the LDB.  The
observational error along with the observed LDB luminosity for each
cluster is shown as the labeled crosses in Figure~\ref{errfig}.

We then calculate the LDB age and errors that result from
observational, theoretical, and bolometric correction uncertainties
using our reference calculation and error models.  The total
percentage age error is a sum in quadrature of these three error
sources.  We derive a LDB age of 148 $\pm$ 19 Myr for the Pleiades
open cluster.  Thus, the LDB age technique rules out ages younger than
91 Myr.  For the Pleiades, the difference in the LDB age and the upper
main-sequence-fitting age without convective-core overshoot (70-80 Myr
- \citet{mer81}) is highly significant.  Even a maximally plausible
0.30 magnitude change in the I-band bolometric correction (consistent
with reconciling the theoretical isochrone fits to the observed (V-I)
color of the Pleiades) would result in an age of 126 $\pm$ 11 Myr for
the Pleiades, where the error includes the observational and
theoretical uncertainty only (see Table~\ref{obstab}).  Alternative
ages and errors of the other open clusters adopting the 0.3 magnitude
change in the I-band bolometric correction are given in parenthesis in
Table~\ref{obstab}.

Our LDB age for the Pleiades is even older than the original LDB age
estimate of 125 Myr given in \citet{sta98}.  In deriving the Pleiades
age, \citet{sta98} employ the calculations of \citet{bar98}.  As
discussed in Sections~\ref{interrbudget} and \ref{exterrbudget}, the
lithium depletion calculations of \citet{bar98} at the lowest
luminosities/oldest ages are younger by 15\%.  The 15\% younger age is
a 5-$\sigma$ deviation from our reference model based on our
examination of the theoretical errors in the LDB ages.  Using
identical input physics as close as possible to \citet{bar98}, we are
unable to reproduce their young LDB ages at the lowest luminosities.
Until other independent calculations are available, including an
additional 5\% error to the LDB ages may be warranted.  This
systematic error is discussed in Section~\ref{exterrbudget} and is not
included in our error budget as it does not materially affect any
conclusions.

For the Pleiades cluster, the LDB age technique is
dominated by uncertainties in the bolometric correction and the
observational LDB location.  If the bolometric correction and LDB
location error sources are eliminated, then our calculations show the
LDB age technique could provide absolute ages for open clusters
accurate to 3\% for 200 Myr clusters increasing to 8\% for 20 Myr
clusters.

\acknowledgements We want to thank I. Baraffe, J. Stauffer, and
D. Terndrup for useful discussions.  This work was supported by NSF
grant AST-0206008.

\begin{deluxetable}{ccccccccc}
\rotate
\tablewidth{0pt}
\tablecaption{Reference Li Depletion Data\tablenotemark{a}\label{reftab}}
\tablehead{\colhead{$M/M_\odot$} & \colhead{Age} & \colhead{Log(L/L$_{\odot}$)} & \colhead{Log($R/R_\odot$)} & \colhead{Log(g)} & \colhead{Log(Teff)} & \colhead{$\sigma_{THY}$} & \colhead{$\sigma_{BC}$} & \colhead{M$_{I}$} \\ \colhead{} & \colhead{(Myr)} & \colhead{} & \colhead{} & \colhead{} & \colhead{} & \colhead{(\% Age)} & \colhead{(\% Age)} & \colhead{}}
\startdata
0.30  & 18.08 & -1.39 & -0.253 & 4.42 & 3.54 &  8.0  & 6.0 & 7.84 \\
0.25  & 22.53 & -1.58 & -0.321 & 4.48 & 3.53 &  7.7  & 6.0 & 8.37 \\
0.20  & 29.67 & -1.80 & -0.406 & 4.55 & 3.52 &  7.0  & 6.0 & 8.98 \\
0.175 & 34.76 & -1.92 & -0.458 & 4.60 & 3.51 &  6.6  & 6.0 & 9.30 \\
0.15  & 41.53 & -2.05 & -0.519 & 4.65 & 3.51 &  6.2  & 6.0 & 9.64 \\
0.125 & 51.37 & -2.22 & -0.593 & 4.72 & 3.50 &  5.6  & 6.0 & 10.10 \\
0.10  & 68.88 & -2.44 & -0.689 & 4.82 & 3.50 &  4.9  & 6.0 & 10.70 \\
0.09  & 80.89 & -2.56 & -0.738 & 4.87 & 3.49 &  4.5  & 6.0 & 11.04 \\
0.08  & 99.62 & -2.70 & -0.796 & 4.93 & 3.48 &  4.0  & 7.0 & 11.45 \\
0.07  & 139.6 & -2.91 & -0.871 & 5.03 & 3.47 &  3.3  & 9.1 & 12.09 \\
0.065 & 188.8 & -3.08 & -0.924 & 5.10 & 3.45 &  3.0  & 10.8 & 12.64 \\
\enddata
\tablenotetext{a}{Li depleted by a factor of 100 from initial value}
\end{deluxetable}

\begin{deluxetable}{ccccccccccccccc}
\rotate
\tablewidth{0pt}
\tablecaption{Observed Cluster Properties\label{obstab}}
\tablehead{\colhead{Cluster} & \colhead{I$_{LDB}$} & \colhead{$\sigma_{LDB}$} & \colhead{$(m-M)_{o}$} & \colhead{$\sigma_{DM}$} & \colhead{$A_{I}$} & \colhead{$\sigma_{A}$} & \colhead{$M_{I}$} & \colhead{$\sigma_{M}$} & \colhead{Age\tablenotemark{a}} & \colhead{$\sigma_{OBS}$} & \colhead{$\sigma_{THY}$} & \colhead{$\sigma_{BC}$} & \colhead{$\sigma_{TOT}$} & \colhead{$\sigma_{AGE}$\tablenotemark{b}} \\ \colhead{} & \colhead{} & \colhead{} & \colhead{} & \colhead{} & \colhead{} & \colhead{} & \colhead{} & \colhead{} & \colhead{(Myr)} & \colhead{(\% Age)} & \colhead{(\% Age)} & \colhead{(\% Age)} & \colhead{(\% Age)} & \colhead{(Myr)}}
\startdata
Pleiades   & 17.86 & 0.10 & 5.60 & 0.10 & 0.06 & 0.03 & 12.20 & 0.14 & 148 (126) &  8.1 &  3.2 & 9.4 & 13 &  19 (11)\\
$\alpha$ Per & 17.87 & 0.15 & 6.23 & 0.10 & 0.17 & 0.04 & 11.47 & 0.18 & 101 (87) & 8.9 &  4.0 & 7.1 & 12 & 12 (9)\\
IC 2391    & 16.22 & 0.15 & 5.95 & 0.10 & 0.02 & 0.02 & 10.25 & 0.18 &  55 (48) &  8.1 &  5.4 & 6.0 & 11 &  6 (5)\\
NGC 2547   & 18.00 & 0.25 & 8.15 & 0.15 & 0.05 & 0.03 &  9.80 & 0.29 &  45 (38) & 13.3 &  6.0 & 6.0 & 16 &  7 (7)\\
\enddata
\tablenotetext{a}{Value in parenthesis is the LDB age for a -0.3 magnitude shift in the I-band bolometric correction (see Section~\ref{bolosec}).}
\tablenotetext{b}{Value in parenthesis is the error in the LDB age for a -0.3 magnitude shift in the I-band bolometric correction.  This error includes observational and theoretical errors only.}
\end{deluxetable}

\end{document}